\documentclass[preprint]{aastex}

\usepackage{emulateapj5}
\usepackage{apjfonts}

\renewcommand\tablecaption[1]{\centerline{\sc #1}\vskip 6pt}
\renewcommand\colhead[1]{\normalsize{#1}}
\renewenvironment{deluxetable}{\begin{tabular}}{\end{tabular}}

\newcommand{\cago}{$^{12}{\rm C}(\alpha,\gamma)^{16}{\rm O}$\ }

\shorttitle{MEASURING \cago FROM ASTEROSEISMOLOGY}

\shortauthors{METCALFE, SALARIS, \& WINGET}

\begin{document}

\title{Measuring \cago from White Dwarf Asteroseismology}

\author{T.S. Metcalfe\altaffilmark{1}, M. Salaris\altaffilmark{2},
and D.E. Winget\altaffilmark{3}}

\altaffiltext{1}{Theoretical Astrophysics Center, Institute of Physics 
and Astronomy, Aarhus University, 8000 Aarhus C, Denmark}

\altaffiltext{2}{Astrophysics Research Institute, Liverpool John Moores 
University, Twelve Quays House, Egerton Wharf, Birkenhead, CH41 1LD, UK}

\altaffiltext{3}{Department of Astronomy, Mail Code C1400, University of 
Texas, Austin, TX 78712, USA}

\email{<travis@ifa.au.dk>\ <ms@astro.livjm.ac.uk>\ <dew@astro.as.utexas.edu>}

\submitted{Accepted to The Astrophysical Journal}

\begin{abstract}

During helium burning in the core of a red giant, the relative rates of
the 3$\alpha$ and \cago reactions largely determine the final ratio of
carbon to oxygen in the resulting white dwarf star. The uncertainty in the
3$\alpha$ reaction at stellar energies due to the extrapolation from
high-energy laboratory measurements is relatively small, but this is not
the case for the \cago reaction. Recent advances in the analysis of
asteroseismological data on pulsating white dwarf stars now make it
possible to obtain precise measurements of the central ratio of carbon 
to oxygen, providing a more direct way to measure the \cago reaction rate 
at stellar energies. We assess the systematic uncertainties of this 
approach and quantify small shifts in the measured central oxygen 
abundance originating from the observations and from model settings 
that are kept fixed during the optimization. Using new calculations of 
white dwarf internal chemical profiles, we find a rate for the \cago 
reaction that is significantly higher than most published values. The 
accuracy of this method may improve as we modify some of the details of
our description of white dwarf interiors that were not accessible 
through previous model-fitting methods.

\end{abstract}


\section{INTRODUCTION}

The nuclear reactions responsible for the synthesis of intermediate-mass
elements in stars have been known for more than four decades. Although no
existing particle accelerator can measure cross-sections for any of these
reactions at the low energies relevant to stellar interiors, theoretical
methods have been devised to extrapolate from experimental measurements
obtained at higher energies. We can only test the validity of these
extrapolations indirectly, by predicting the astronomical consequences of
a given set of nuclear reaction rates, and then comparing the predictions
to observations.

By necessity the astronomical predictions are themselves model-dependent.
For example, models of stellar evolution with detailed compilations of
nuclear reaction rates have been used to predict the nucleosynthetic
yields of various elements, and these were compared to the observed solar
abundances to test the accuracy of the rates \citep{ww93}. Although this
experiment relies on some processes that are not entirely understood (star
formation rates, convective mixing and overshoot, supernova explosion
mechanisms), it has provided some of the most convincing evidence to date
that the extrapolated rates are reasonably accurate.

Such an analysis is simplified by the fact that most of the relevant
nuclear reactions have a strong temperature dependence, so large
uncertainties in the absolute rates correspond to very small changes to
the temperature in models of the stellar interior. The notable exception
is the \cago reaction, and so it has been the focus of many experiments in
both astronomy and physics.

During helium burning in the core of a red giant, the 3$\alpha$ and \cago
reactions compete for the available helium nuclei. The relative rates of
the two reactions determines the final yield of oxygen deep in the core.  
The 3$\alpha$ rate is well established, but the same is not true of the
\cago reaction. The extrapolation of its rate to stellar energies is
complicated by interference between various contributions to the total
cross-section, leading to a relatively large uncertainty
\citep[cf.][]{kun02}. This translates into similarly large uncertainties
in our understanding of every astrophysical process that depends on this
reaction, from supernovae explosions to galactic chemical evolution.

\section{METHOD}

\cite{ww93} derived a measurement of the \cago cross-section based on the
fact that the total oxygen production in high mass stars has a dramatic
effect on the yields of many other elements with masses between oxygen and
iron. The end-result of any adopted value for the \cago cross-section is a
distinct set of abundances produced and distributed during the explosion
of the stellar models as supernovae. The optimal value is the one that can
most closely reproduce the relative abundances observed in the Sun.

Another branch of stellar evolution provides an independent method of
probing the composition of the material in the stellar interior after
helium burning. A star that is not massive enough to establish an internal
temperature and density sufficient to fuse two carbon nuclei will
eventually shed its envelope as a planetary nebula, leaving the exposed
core to begin a long life as a white dwarf star. Depending on the
composition of its stratified surface layers, an otherwise normal white
dwarf will begin pulsating as it cools through one of several distinct
ranges of temperature, or instability strips. During this phase,
astronomers have an opportunity to probe the stellar interior while it is
still intact, using the techniques of asteroseismology.

In the past decade, the observational requirements of white dwarf
asteroseismology have been satisfied by the development of the Whole Earth
Telescope \citep[WET;][]{nat90}, a group of astronomers distributed around
the globe who cooperate to observe these stars continuously for up to two
weeks at a time. This instrument has now provided a wealth of
seismological data on the different varieties of pulsating white dwarf
stars \citep[cf.][] {win91,win94,kle98}.

In an effort to bring the analysis of WET data to the level of
sophistication demanded by the observations, \cite{mnw00} recently
developed a new fitting method using a genetic algorithm to determine the
optimal model parameters, including the stellar mass ($M_*$), the
effective temperature ($T_{\rm eff}$), and the mass of the surface helium
layer ($M_{\rm He}$). The underlying ideas for genetic algorithms were
inspired by Charles Darwin's notion of biological evolution through
natural selection. The basic idea is to solve a problem by {\it evolving}
the best solution from an initial set of random guesses. The computer
model provides the framework within which the evolution takes place, and
the individual parameters controlling it serve as the genetic building
blocks. Observations provide the selection pressure. In practice, this
method can be much more efficient than other comparably global techniques,
and provides an objective determination of the optimal set of model
parameters along with some indication of the uniqueness.

Using this new approach \citeauthor{mwc01} (\citeyear[][hereafter
MWC]{mwc01}) derived an optimal value for the central oxygen mass fraction
in the pulsating white dwarf GD~358 using a simple parameterization of the
internal oxygen profile. This parameterization fixed the oxygen mass
fraction to its central value ($X_{\rm O}$) out to some fractional mass
($q$) where it then decreased linearly in mass to zero oxygen at 0.95
$m/M_*$. The statistical uncertainties on these two parameters were
determined by calculating a grid of 10,000 models forming all combinations
of $X_{\rm O}$ and $q$, with the other three parameters fixed at their
optimal values since none of them showed any correlation with $X_{\rm O}$.  
The internal 1$\sigma$ uncertainties were then defined by the range of
models with root-mean-square (rms) residuals that differed from the
optimal solution by less than the statistical uncertainties on the
observed periods \citep[quantified by][]{mnw00}. MWC then derived a
preliminary value of the astrophysical S-factor at 300 keV ($S^{\rm
MWC}_{300}=290\pm15$ keV b)  for the \cago reaction by extrapolating the
calculations of \cite{sal97} to the mass and oxygen abundance of the
optimal model for GD~358 ($X_{\rm O}=0.84\pm0.03$).

The sensitivity of our pulsation models to the core composition arises 
through differences in the Brunt-V\"ais\"al\"a (BV) frequency due to 
changes in the mean molecular weight of the internal mix of ions. For the 
purposes of illustration, consider the idealized case where the electrons 
can be treated as a completely degenerate gas and the ions as an ideal gas. 
As is known, the electron degeneracy pressure is much larger than that due
to the ions, and at a given density it is independent of the composition 
(C and O have the same number of electrons per nucleon). The only non-zero 
contribution to the BV frequency ($N^2$) must therefore come from the 
partial pressure of the ions, and for an ideal gas this leads to $N^2\propto
\rho T/\mu_{\rm ion}$, where $\mu_{\rm ion}$ is the mean molecular weight 
of the ions. Consequently, the BV frequency is lower in O-core models 
($\mu_{\rm ion}=16$) relative to C-core models ($\mu_{\rm ion}=12$).  
Of course, the electrons are not completely degenerate and also contribute 
a non-negligible amount to the BV frequency, which tends to reduce the
magnitude of this effect. For parameters similar to the optimal model for 
GD~358, O-core models have a BV frequency that is lower by $\sim$10\% in 
the inner 95\% of the stellar mass compared to C-core models. Since this 
influences such a large fraction of the mass, the net effect on the periods 
is nearly as significant as the effect of the envelope He/C interface 
\citep[cf.][Fig.4]{mmw01}.

\section{SYSTEMATIC UNCERTAINTIES}

In this paper we examine the various contributions to the systematic
uncertainty on the measurement of $X_{\rm O}$, including those originating
from the observations themselves and from the theoretical models used to
interpret them. The latter category includes contributions from the
settings of the white dwarf evolution/pulsation models, and from the
models used to produce new white dwarf internal chemical profiles
\cite[following][]{sal97} that match the derived value of $X_{\rm O}$.

\subsection{Observational Data}

GD~358 has been the target of three coordinated observing campaigns by the
WET. The first campaign occurred in 1990, and the results \citep[reported
in][]{win94} were the observational basis for the global model-fitting of
MWC because more pulsation modes were observed in 1990 than during the
other two campaigns. The results of the second campaign in 1994 were
reported in \cite{vui00}, and the data from the third campaign in 2000 are
still undergoing analysis. Although the power spectra resulting from the
three campaigns differ substantially---with modes changing their relative
amplitudes and sometimes falling below the sensitivity of the
observations---the underlying normal mode structure appears to be
remarkably stable (see Table \ref{tab1}).

To assess the systematic uncertainty on the optimal model parameters due
to the observational data, we repeated the global model-fitting procedure
of MWC using the periods and period-spacings from the other two campaigns
on GD~358. For the 1994 data we used the $\ell=1, m=0$ modes listed in
Table 3 of \cite{vui00} together with the 1990 values for the $k=10,11,12$
modes, which had no reliable identification in the second campaign.
Despite some significant differences in the power spectrum of the third
campaign, most of the $\ell=1, m=0$ modes present in 1990 were also
detected in 2000, which we combined with the 1990 values for $k=11,12,15$.
Although the three data sets are not entirely independent, this procedure
should provide some information about the systematic uncertainties due to
subtle differences in the observational data.

The differences between the optimal set of model parameters from MWC
($T_{\rm eff}=22,600~{\rm K}$, $M_*=0.650~M_{\sun}$, $\log[M_{\rm
He}/M_*]=-2.74$, $X_{\rm O}=0.84$, $q=0.49$) and the results of new fits
to these two additional data sets are shown in Table \ref{tab2} [Fits $a$
and $b$]. Also listed are differences for a new fit to the 1990 data
supplemented by the identifications for $k=19,20,21$ from \cite{bra93}
[Fit $c$], and for a subset of the 1990 data using observations from only
one telescope site [McDonald Observatory, Fit $d$]. The latter fit
demonstrates that a multi-site campaign is necessary for an accurate
determination of the internal composition from ground-based observations.
For each fit we also list the rms differences between the observed and
calculated periods ($\sigma_P$) and period spacings ($\sigma_{\Delta P}$).
For comparison, the optimal fit from MWC had $\sigma_P=1.28\ s$ and
$\sigma_{\Delta P}=1.42\ s$.

\subsection{Pulsation Model Settings}

There are a number of settings in the evolution/pulsation models used by
MWC that were held constant during the optimization procedure. These fixed
settings included: [1] the fractional mass of the boundary between the
fully self-consistent core and the envelope of the models ($q_{\rm env}$),
[2] the fractional mass where the oxygen abundance is fixed to zero
($q_0$), [3] the 

\begin{table*}
\begin{center}
\vspace*{12pt}
\tablenum{1}
\centerline{\sc Table \ref{tab1}}
\tablecaption{Observational data sets for GD~358 \label{tab1}}
\begin{deluxetable}{lcccc}
\hline\hline
\colhead{k} & 
\colhead{WET 1990$^{1,2}$} & 
\colhead{WET 1994$^{3}$} & 
\colhead{WET 2000\tablenotemark{a}} &
\colhead{McD 1990}  
\\ \hline
8$\dotfill$ & 423.27  & 423.26                  & 423.24                  & 422.58 \\
9$\dotfill$ & 464.23  & 464.23                  & 464.25                  & 464.25 \\
10$\ldots$  & 501.59  & 501.59\tablenotemark{b} & 501.20                  & 502.50 \\
11$\ldots$  & 541.75  & 541.75\tablenotemark{b} & 541.75\tablenotemark{b} & 538.31 \\
12$\ldots$  & 576.76  & 576.76\tablenotemark{b} & 576.76\tablenotemark{b} & 571.31 \\
13$\ldots$  & 618.28  & 617.88                  & 618.44                  & 616.11 \\
14$\ldots$  & 658.35  & 657.91                  & 658.62                  & 657.11 \\
15$\ldots$  & 700.64  & 701.06                  & 700.64\tablenotemark{b} & 700.62 \\
16$\ldots$  & 734.30  & 733.96                  & 734.19                  & 736.20 \\
17$\ldots$  & 770.67  & 770.58                  & 771.25                  & 770.74 \\
18$\ldots$  & 810.7~\ & 809.37                  & 811.19                  & 808.63 \\
19$\ldots$  & 854.8\tablenotemark{c}~\ &    -   &    -                    &    -   \\
20$\ldots$  & 894.2\tablenotemark{c}~\ &    -   &    -                    &    -   \\
21$\ldots$  & 943.5\tablenotemark{c}~\ &    -   &    -                    &    -   \\
\hline\hline
\end{deluxetable}
\end{center}
\vspace*{-6pt}

\hskip 1.85in {$^a$}{Results of preliminary data analysis.}

\hskip 1.85in {$^b$}{Adopted value from WET 1990.}

\hskip 1.85in {$^c$}{Included for Fit $c$ (see Table \ref{tab2}).}

\hskip 1.85in {\footnotesize {\sc References}---(1) \cite{win94}; (2) \cite{bra93}; (3) \cite{vui00}}

\end{table*}


\noindent prescription for convection (fixed to ML3), [4] the
diffusion coefficients that describe the He/C interface in the envelope
(fixed at $\alpha_{3,4}=\pm3$), and [5] the energy losses due to neutrinos
($E_\nu$).

We determined the magnitude of systematic uncertainties due to these fixed
settings by repeating the 5-parameter global model-fitting procedure of
MWC with the settings fixed at alternate values. Anticipating that the
shifts to the optimal model parameters would be relatively small, we were
able to complete the calculations in fewer iterations than MWC without
sacrificing the accuracy of the final solution.

After changing one of the fixed model settings, we fixed $q$ at its value
from MWC (0.49 $m/M_*$) and performed 10 runs of the genetic algorithm. We
then temporarily fixed $T_{\rm eff}$, $M_*$, and $M_{\rm He}$ to their
optimal values from this first iteration and computed a small grid of
models with a range of $X_{\rm O}$ and $q$ around their optimal values. If
a better fit existed in this grid with a different value of $q$, we
performed a new set of genetic algorithm runs with $q$ fixed at this
better value and repeated the procedure until no better value of $q$ was
found in the small grid.  Finally, we fixed the mass to the optimal value
from the final iteration with fixed-$q$ and verified that the optimal
fixed-mass solution from 10 more runs of the genetic algorithm was
identical to that found in the fixed-$q$ case.

Since the final fixed-mass and fixed-$q$ iterations were completely
independent, the end result should be just as global as the solution found
by MWC. Even with this computational shortcut, the cumulative additional
model-fitting performed for this study represents more than 5
GHz-CPU-years of calculation, which was only practical because of the
dedicated parallel computers available to this project
\citep[cf.][]{mn00}.

Table \ref{tab3} lists the changes to the optimal set of parameters for
the new 5-parameter fits with seven variations to the fixed model
settings, along with the corresponding values of $\sigma_P$ and
$\sigma_{\Delta P}$. We discuss the details of these changes below.

\subsubsection{Core-Envelope Boundary}

As noted by \cite{mnw00} the evolutionary calculations for the core of our
models are fully self-consistent, but the envelope is treated separately
and adjusted to match the boundary conditions at each time step. As a
consequence, only the fraction of the mass inside the core/envelope
boundary is directly involved in the evolution, and this may lead to
systematic errors. MWC used a grid of starter models with this boundary
fixed at $q_{\rm env}=0.95~m/M_*$. Fit $e$ in Table \ref{tab3} is the
result of using a new grid of starter models with this boundary fixed
further out, at $q_{\rm env}=0.98~m/M_*$.

\subsubsection{Zero Oxygen Point}

The parameterization of the internal oxygen profiles used by MWC fixed the
fractional mass where the oxygen abundance goes to zero at
$q_0=0.95~m/M_*$. They did this because the models cannot presently
include oxygen in the envelopes, leading to the constraint that $q_0 \le
q_{\rm env}$. The new grid of starter models with $q_{\rm env}=0.98~m/M_*$
allowed us to move $q_0$ out further. Fit $f$ in Table \ref{tab3} shows
the shift relative to Fit $e$ when $q_0$ is fixed at 0.98 $m/M_*$.

\subsubsection{Mixing-Length Prescription \label{mlsec}}

Our evolution/pulsation code can use either the ML1 prescription for
convection from \cite{boh58}, or the prescription of \cite{bc71} with the
mixing-length/pressure scale height ratio set to either 1 or 2, which
correspond to ML2 and ML3 respectively. \cite{bea99} found that the best
internal consistency between optical and IUE-based effective temperature
determinations for DB stars can be achieved using a mixing-length/pressure
scale height ratio between the standard ML2 and ML3. The optimal model
found by MWC assumed ML3 convection. Fits $g$ and $h$ in Table \ref{tab3}
show the shifts to the optimal parameter values when ML2 or ML1 are used
respectively.

\begin{table*}
\begin{center}
\vspace*{12pt}
\tablenum{2}
\centerline{\sc Table \ref{tab2}}
\tablecaption{Systematic differences from various data sets\label{tab2}}
\begin{deluxetable}{clccccccc}
\hline\hline
\colhead{Fit} & 
\colhead{Data Set} & 
\colhead{$\Delta T_{\rm eff}$} & 
\colhead{$\Delta M_*/M_{\odot}$} &
\colhead{$\Delta \log(M_{\rm He}/M_*)$} & 
\colhead{$\Delta X_{\rm O}$} & 
\colhead{$\Delta q$} &
\colhead{$\sigma_P$} &
\colhead{$\sigma_{\Delta P}$}
\\ \hline
$a\ldots$ & 1994 run    &     -  & - & - & $+$0.02 &     -   & 1.31 & 1.37 \\
$b\ldots$ & 2000 run    &     -  & - & - & $-$0.05 & $-$0.01 & 1.23 & 1.49 \\
$c\ldots$ & extra-k     & $-$100 & - & - & $+$0.08 &     -   & 2.28 & 2.11 \\
$d\ldots$ & single-site &     -  & - & - & $+$0.14 & $+$0.01 & 1.86 & 1.50 \\
\hline\hline
\end{deluxetable}
\end{center}
\vspace*{-12pt}
\end{table*}


\begin{table*}
\begin{center}
\vspace*{12pt}
\tablenum{3}
\centerline{\sc Table \ref{tab3}}
\tablecaption{Systematic differences from pulsation model settings\label{tab3}}
\begin{deluxetable}{clccccccc}
\hline\hline
\colhead{Fit} & 
\colhead{Setting} & 
\colhead{$\Delta T_{\rm eff}$} & 
\colhead{$\Delta M_*/M_{\odot}$} &
\colhead{$\Delta \log(M_{\rm He}/M_*)$} & 
\colhead{$\Delta X_{\rm O}$} & 
\colhead{$\Delta q$} &
\colhead{$\sigma_P$} &
\colhead{$\sigma_{\Delta P}$}
\\ \hline
$e\ldots$ & $q_{\rm env}$        & $+$1200 & $-$0.030 & $+$0.10 & $+$0.11 & $-$0.01 & 1.46 & 1.21 \\
$f\ldots$ & $q_{\rm 0}$          &     -   &    -     &   -     & $+$0.04 &   -     & 1.47 & 1.18 \\
$g\ldots$ & ML2                  &  $+$300 & $-$0.010 & $+$0.05 & $+$0.06 &   -     & 1.27 & 1.48 \\
$h\ldots$ & ML1                  &  $+$300 & $-$0.010 & $+$0.05 & $+$0.06 &   -     & 1.27 & 1.46 \\
$i\ldots$ & $\alpha_{3,4}(\pm2)$ &  $-$100 &    -     & $+$0.05 & $-$0.02 &   -     & 1.46 & 1.45 \\
$j\ldots$ & $\alpha_{3,4}(\pm4)$ &  $+$100 &    -     &   -     & $-$0.02 & $-$0.01 & 1.26 & 1.58 \\
$k\ldots$ & $E_\nu$              &     -   &    -     &   -     & $-$0.05 & $-$0.01 & 1.27 & 1.42 \\
\hline\hline
\end{deluxetable}
\end{center}
\vspace*{-12pt}
\end{table*}


\subsubsection{Diffusion Coefficients \label{dcsec}}

The description of the He/C transition in our model envelopes does not
include the effects of time-dependent diffusion directly, as is done by 
\cite{dk95} and \cite{cor02}. Rather, we use a parameterization of the
method of \cite{af80} using the trace element approximation with two
adjustable diffusion coefficients: one to describe the shape of the
profile in the carbon-rich region of the envelope, and the other to
describe the shape in the helium-rich region \citep[cf.][]{mw00,tfw90}.
The equilibrium values for these coefficients are
$\alpha_{3,4}=^{+2}_{-2/3}$, but the absolute values of each of them may
be larger, depending on the relative magnitude of the evolutionary and
diffusive timescales for a specific model. For the purposes of
asteroseismology, these coefficients can be considered free parameters. By
choosing coefficients of opposite sign but the same absolute value, we
ensure a smooth composition transition profile. This avoids the unphysical
mode-trapping features introduced by the equilibrium diffusion values, but
still provides a reasonable approximation to the expected profile shape
\citep[cf.][Fig.~2]{bww93}. The diffusion coefficients used by MWC were
$\alpha_{3,4}=\pm3$. Fits $i$ and $j$ in Table \ref{tab3} show the effect
of using coefficients that produce thicker and thinner He/C transition
zones, corresponding to $\alpha_{3,4}=\pm2$ and $\alpha_{3,4}=\pm4$
respectively.

\subsubsection{Neutrino Losses}

The energy losses from neutrino emission can alter the thermal structure 
of our models slightly, since they cool the deep interior more efficiently 
than the outer regions of the core. For the optimal model of GD~358 
(which has $\log T_{\rm c}=7.52$), the neutrino luminosity still amounts 
to $\sim$20\% of the total luminosity. The effect on the internal 
temperature is small, with a maximum difference of $\Delta T/T=0.25\%$ in 
the core. At some level, this might have a detectable signature on the 
pulsation properties of the models. The optimal solution found by MWC 
relied on the neutrino rates of \cite{ito96}. Fit $k$ in Table \ref{tab3} 
shows the result of ignoring all sources of energy loss due to neutrinos 
during the white dwarf cooling phase.

\subsection{Internal Chemical Profile Model Settings}

As noted by MWC, converting a measurement of the central oxygen abundance
into a more precise constraint on the \cago reaction rate requires new
evolutionary calculations. To produce profiles that match the mass and
central oxygen abundance of the optimal fit to GD~358, we have used the
same method and code described in \cite{sal97}, but updated to use the
nuclear reaction rates from the NACRE collaboration \citep{ang99} rather
than from \cite{cau85}.

When a white dwarf is being formed in the core of a red giant during
helium burning, the 3$\alpha$ and \cago reactions 
are competing for the
available helium nuclei. The relative success of the two reactions is the
primary factor in determining the final oxygen 
abundance in the center of a white dwarf star 
formed in this way. A secondary factor is the
efficiency and extent of mixing. The profile calculations normally include
the effects of semiconvection during the central helium burning phase, but
optionally allow complete mixing in the convective overshooting region.
This is followed by Rayleigh-Taylor re-homogenization prior to the white
dwarf cooling phase, which generally redistributes inward the extra oxygen
that accumulates near the outer boundary of the central convective region
\citep[see][and references therein]{sal97},

In this section we quantify the magnitude of systematic uncertainties on
the central oxygen abundance in our models of the internal chemical
profiles from changes to these and other aspects of the input physics.
Figures \ref{fig1} and \ref{fig2} show the internal oxygen profiles for
0.61 $M_\sun$ models resulting from the changes we have considered. We
discuss the details of these changes below.

\subsubsection{Nuclear Reaction Rates}

The top two panels of Figure \ref{fig1} show the effect of the statistical
uncertainties on the NACRE reaction rates for the \cago and 3$\alpha$
reactions. In both cases the solid line is the result of using the
suggested values for the reaction rates. In\hfill Figure \ref{fig1}a,\hfill 
the\hfill dotted\hfill and\hfill dashed\hfill lines\hfill indicate\hfill 
the\hfill profiles

\epsfxsize 3.5in
\epsffile{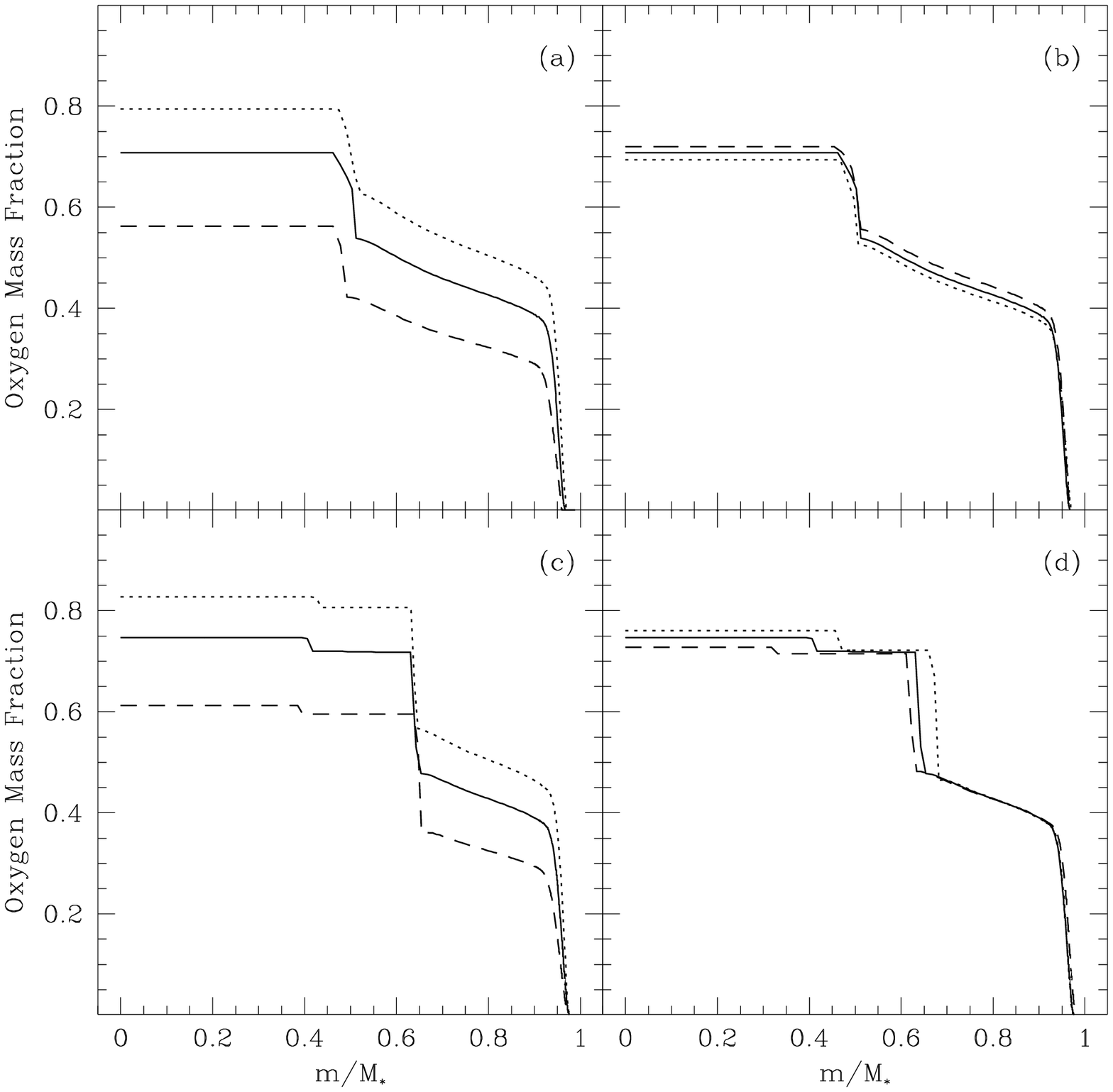}
\figcaption[f1.eps]{Internal oxygen profiles for a $0.61~M_\sun$ white
dwarf model using the standard NACRE rates (solid), and (a) using the
upper (dotted) and lower (dashed) limits for the $^{12}{\rm
C}(\alpha,\gamma)^{16}{\rm O}$ rate, (b) using a value for the 3$\alpha$
rate 10 percent higher (dotted) and lower (dashed) than the NACRE value,
(c) as in (a) but including convective overshoot with $\alpha_{\rm
ov}/H_p=0.20$, (d) including convective overshoot with $\alpha_{\rm
ov}/H_p=0.25$ (dotted) and $\alpha_{\rm ov}/H_p=0.15$
(dashed).\label{fig1}}
\vskip 12pt

\noindent that result when using the upper
and lower limits respectively for the \cago reaction, yielding values for
the central oxygen abundance between 0.56 and 0.79 ($\Delta X_{\rm
O}=^{+0.08}_{-0.15}$).

In Figure \ref{fig1}b, the dotted and dashed lines are the profiles
resulting from a value for the 3$\alpha$ rate that is 10 percent higher
and lower than the recommended value respectively. This leads to a range
of central oxygen abundances from 0.69 to 0.72 ($\Delta X_{\rm
O}=^{+0.01}_{-0.02}$).

\subsubsection{Convective Overshoot}

The bottom two panels of Figure \ref{fig1} show profiles that include the
effects of convective overshoot with complete mixing in the overshooting
region during central helium burning, rather than standard semiconvective 
mixing. The profiles in Figure
\ref{fig1}c were produced in a similar manner as the corresponding curves
in Figure \ref{fig1}a, but include convective overshoot with the
overshooting parameter fixed at $\alpha_{\rm ov}=0.20~H_p$, as suggested
by the recent analysis of eclipsing binary data by \cite{rjg00}. With
overshoot included, the uncertainty in the \cago reaction leads to values
for the central oxygen abundance from 0.61 to 0.82 ($\Delta X_{\rm O}=
^{+0.08}_{-0.13}$). The net effect of including convective overshoot on
average is to increase the central oxygen abundance at a given \cago rate
by $\Delta X_{\rm O}=+0.04$.

In Figure \ref{fig1}d, the dotted and dashed lines are the profiles from
higher ($\alpha_{\rm ov}=0.25~H_p$) and lower ($\alpha_{\rm ov}=0.15~H_p$)
assumed values for the overshooting parameter respectively. The central
oxygen abundance ranges from 0.73 to 0.76 ($\Delta X_{\rm O}= ^{+0.01}
_{-0.02}$).

\subsubsection{Equation of State \& Metallicity}

In Figure \ref{fig2} we show the profiles resulting from modifications to
two other aspects of the input physics. To produce the dashed line,\hfill we
\hfill computed\hfill the\hfill progenitor\hfill evolution\hfill after 
\hfill making\hfill an 

\epsfxsize 3.5in
\epsffile{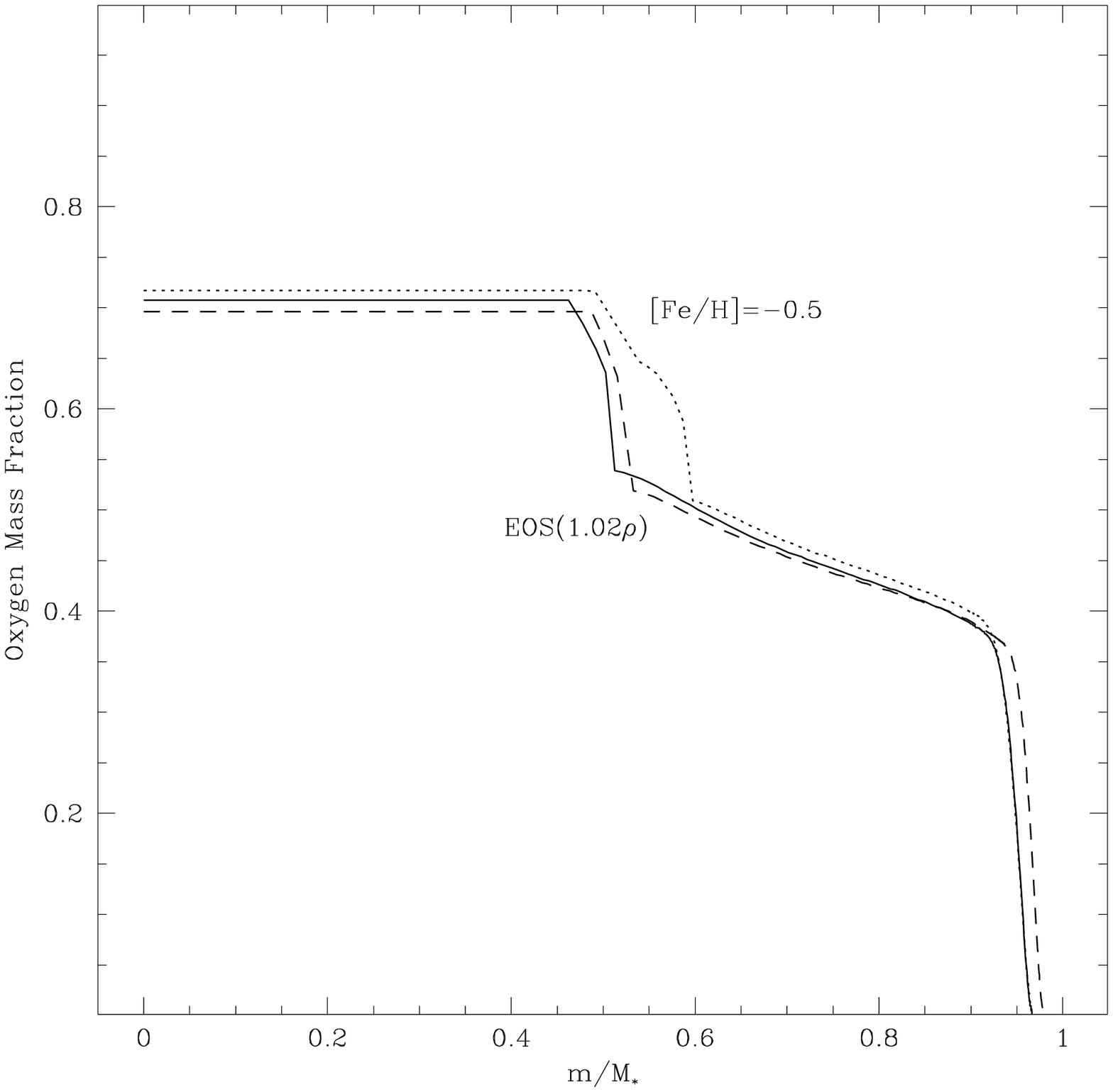}
\figcaption[f2.eps]{The internal oxygen profiles for a $0.61~M_\sun$ 
white dwarf model using the NACRE rates (solid line), with the density
artificially increased by 2 percent (dashed line), and from a progenitor
with a metallicity of $[{\rm Fe/H}]=-0.5$ (dotted line).\label{fig2}}
\vskip 12pt

\noindent {\it ad hoc}
modification to the equation of state (EOS). At fixed values of $P$ and
$T$ we increased the density by 2 percent. We did not modify the internal
energy (the specific heats were unchanged),
so the modified EOS has
different $P$-$T$-$\rho$ and $\rho$-$U$ relationships. Finally, we changed
the adiabatic gradient to make it consistent with this pressure and
internal energy \citep[cf.][]{kw94}. The change in the central oxygen
abundance due to this modified EOS is $\Delta X_{\rm O}=-0.01$.

The probable age of GD~358 ($\sim0.1$ Gyr) leads us to expect the
metallicity of its progenitor to be approximately solar ($[{\rm
Fe/H}]=0.0$).\hfill For\hfill completeness,\hfill we\hfill tested\hfill 
the\hfill sensitivity 
of the oxygen
profile to changes in the progenitor metallicity. The dotted line in
Figure \ref{fig2} shows the profile from a progenitor with an initial
$[{\rm Fe/H}]=-0.5$. The primary consequence of this change was to extend
the O-rich region, but it also increases the central oxygen abundance by
$\Delta X_{\rm O}<0.01$.

\subsection{Shape of the C/O Profile}

It is clear from the ``reverse approach'' of MWC that the detailed shape
of the internal C/O profile has the potential to reduce the rms residuals
of the optimal fit. It is unclear whether the improvement can be
considered significant, since the method optimizes the shape of the
Brunt-V\"ais\"al\"a frequency directly and is not strictly limited to
physically plausible profiles. The parameterization of the internal
chemical profile we are using for forward modeling has no physical basis,
but was used to facilitate comparison with previous work
\citep[cf.][]{bww93}. It is unclear whether our fits would benefit from
incorporating a profile with a physical basis because there are still
significant differences between the evolutionary profiles computed by
various groups. For example, the evolutionary profiles produced by
\cite{alt01} contain less structure near $q\sim 0.5$ than those of
\cite{sal97}---apparently due to differences in the treatment of the
convective core\hfill boundary\hfill during\hfill helium\hfill burning
\hfill (L.~Althaus,\hfill private

\epsfxsize 3.5in
\epsffile{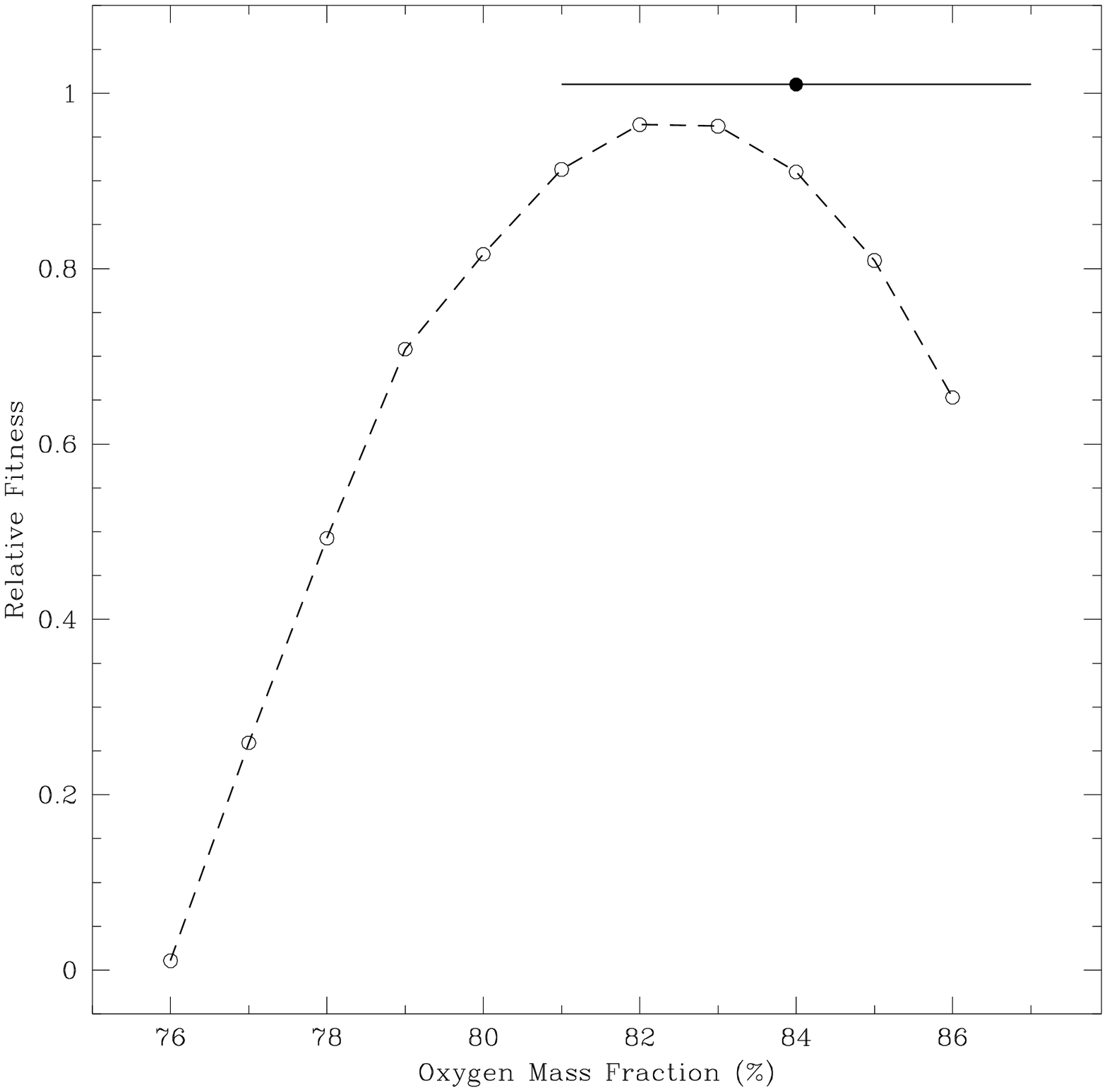}
\figcaption[f3.eps]{The relative quality of the fit to GD~358 using the
optimal set of parameters found by MWC, but using an evolutionary internal
oxygen profile from the calculations of \cite{sal97} scaled to various
values for the central oxygen mass fraction (open points). For comparison
we show the optimal value of the central oxygen mass fraction derived from
the simple parameterization of the internal oxygen profile used by MWC
(solid point; vertical placement is arbitrary).\label{fig3}}
\vskip 12pt

\noindent communication).

To reassure ourselves that the derived values of $X_{\rm O}$ are not
strongly influenced by the use of non-evolutionary internal chemical
profiles, we incorporated the profiles of \cite{sal97} into the optimal
model for GD~358 and scaled it to various values of the central oxygen
abundance. In Figure \ref{fig3} we show the relative quality of the fit
near the peak fitness for these profiles at $X_{\rm O}\sim0.82$. For
reference, we have shown the optimal value of the central oxygen 
abundance derived for GD~358 
from the simple parameterization of the internal oxygen
profile. The {\it absolute} fitness of the optimal model using an
evolutionary profile is much worse, but the peak in the fitness is
consistent with the value derived from the simpler profiles ($\Delta
X_{\rm O} \sim -0.02$). While this is reassuring, a more thorough
exploration of detailed C/O profiles is certainly warranted in the future.

\section{RESULTS}

Having demonstrated that the systematic uncertainties from the
observations and the model settings are small, we adjusted the value of
the \cago rate to match the optimal central oxygen abundance derived for
GD~358 by MWC. In Figure \ref{fig4}, we show profiles for 0.65 $M_\sun$
models using the range of rates for the \cago reaction recommended by the
NACRE collaboration ($S^{\rm NACRE}_{300}=200\pm80$ keV b), and with rates
that produce a central oxygen abundance within the $\pm 1\sigma$ limits
for GD~358. The NACRE rates lead to a range for the central oxygen
abundance between 0.53 to 0.78, with a value of 0.69 at the recommended
rate. To match the central oxygen abundance of GD~358 ($X_{\rm
O}=0.84\pm0.03$) our models require a value for the astrophysical S-factor
at 300 keV of $S_{300}=370\pm40$ keV b, or slightly lower with convective
overshoot included:  $S^{\rm ov}_{300}=360\pm40$ keV b. Note that both the
value and the uncertainty of the ~S-factor ~are larger than the 
~preliminary ~estimate 

\epsfxsize 3.5in
\epsffile{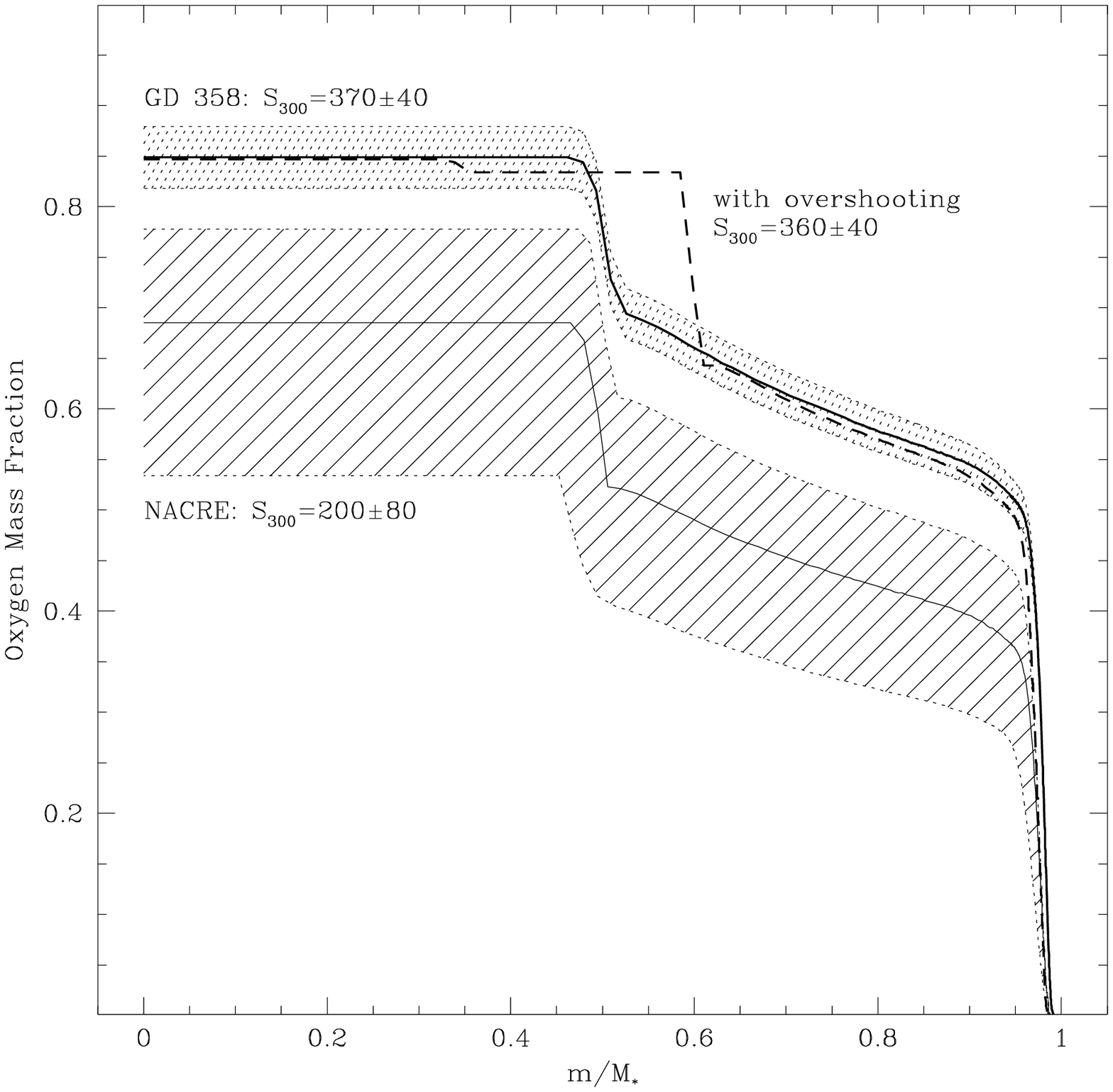}
\figcaption[f4.eps]{The internal oxygen profiles for a $0.65~M_\sun$ white
dwarf model using the NACRE rates (solid line) between the upper and lower
limits on the \cago rate (hashed region). Also shown are the profiles
resulting from the rates that match the central oxygen mass fraction
derived for GD~358 (dark solid line) with the $\pm 1\sigma$ limits (shaded
region), and when overshoot is included (dark dashed line). \label{fig4}}
\vskip 12pt

\noindent 
of MWC, which was based on an extrapolation from the previously
published models of \cite{sal97}.

The oxygen profiles in Figures \ref{fig1} and \ref{fig4} consistently show
the transition from constant oxygen beginning near a fractional mass of
$q\sim0.5$ when the simulations do not include convective overshoot,
regardless of the assumed value for the \cago rate. This is in good
agreement with the value of $q$ found for GD~358 by MWC even though they
did not use evolutionary profiles. With convective overshoot included, the
transition moves out to $q\sim0.6$ and the shape of the profile is
otherwise similar. This leads to the exciting possibility that we can use
asteroseismology to probe the extent of mixing in the stellar interior. If
the progenitor of GD~358 went through normal single-star evolution, then
semiconvective mixing appears to be sufficient to explain the optimal
internal oxygen profile found by MWC. Complete mixing in the overshooting
region is not required.

The ML2 optimal model of GD~358 (Fit $g$ in Table \ref{tab3}) has a mass
and temperature closer to the values inferred from spectroscopy by
\cite{bea99}. This was to be expected since they assumed convective
parameters that are more similar to ML2 than ML3. An independent
observational constraint on the mass and temperature of GD~358 comes from
the luminosity implied by its measured parallax \citep{har85}. The optimal
model from our ML2 fit has $\log(L/L_{\sun})=-1.43$, which is consistent
with the value derived from the parallax, $\log(L/L_{\sun})= -1.32^{+0.17} 
_{-0.27}$ \citep{bw94b}.

An unexpected result from Table \ref{tab3} is the relatively small
difference in the optimal model parameters from changes to the diffusion
coefficients for the He/C transition zone in the envelope. Since the
shifts are only 1 or 2 grid points in each parameter, it may be possible
to optimize the values of the diffusion coefficients using a relatively
small grid search {\it after} finding the global solution with the genetic
algorithm. If the fit can be improved significantly with a different
combination of $\alpha_{3,4}$, we can fix them to their optimal values in
future runs of the genetic algorithm. Eventually, we might begin to
achieve rms residuals that approach the level of the observational
uncertainties ($\sigma_{\rm obs}\sim0.1$ s).

\section{DISCUSSION}

The rate implied for the \cago reaction from GD~358 is high relative to
most published values \citep[cf.][Table 1]{kun02}. The total cross-section 
has electric dipole (E1)
and quadrupole (E2) components, and a variety of experimental and
theoretical methods exist to derive one or both of these contributions.
The highest recently published value for the E1 component is
$S_{300}^{E1}=101\pm17$ keV b \citep{bru01}, while recent work by
\cite{ang01} on the E2 component suggests a possible value in the range
$S_{300}^{E2}=190$--220 keV b. Even if these values ultimately prove to be
correct, the total rate would be in the range $S_{300}=290$--320 keV b,
and would still be only marginally consistent with the value inferred from
GD~358.

Recent work on type Ia supernovae (SNe~Ia) also favors a high value for
the \cago rate to produce model light curves with a sufficiently slow rise
to maximum light \citep{hwt98}. This might even have implications for the
determination of cosmological parameters from high redshift SNe~Ia
\citep{dhs01}. Fortunately, there are other observational consequences of
the higher rate in the infrared spectra of SNe~Ia models, so independent
tests should soon be possible \citep[cf.][]{mhw01}.

If we assume that our method is sound but the value for the \cago rate we
have obtained is not accurate, the simplest explanation is that we have
failed to optimize something in our model, which leads us to a biased
estimate of the central oxygen abundance. In \S 3, we have demonstrated
that many of the fixed model settings do not appear to make much
difference to the final set of optimal parameters. This suggests that we
may want to revisit some of the more fundamental details of our models to
identify the root cause of the discrepancy between our value for the
S-factor and the value extrapolated from high-energy laboratory
measurements. Below we discuss future areas of investigation and possible
improvements to our models that may help us to identify any problems.

The results in Table \ref{tab3} make it clear that the single most
significant improvement we can make to our models is the inclusion of the
entire mass, from core to surface, in the evolution calculations.
Increasing the core mass from 95 to 98 percent of the total stellar mass
changed the optimal value for the central oxygen abundance by 11 percent
and simultaneously resulted in an optimal temperature and stellar mass
closer to the values for GD~358 inferred from spectroscopy \citep{bea99}.
However, the shift in the oxygen abundance from this setting by itself
leads to a {\it higher} inferred nuclear reaction rate, not lower.

It is possible that our method is sound, but we are using a model that is
incomplete. For example, if the interior of GD~358 contains a significant
mass of $^{22}{\rm Ne}$, our fit could be biased toward a higher central
oxygen abundance because it is attempting to create a higher central
density using the wrong ingredients. In fact, we calculated the $^{22}{\rm
Ne}$ abundance along with the other internal chemical profiles, and found
it to be present in the simulations with a mass fraction of $\sim$3-5
percent, depending on the adopted mixing scheme. According to \cite{bh01},
the timescale for gravitational settling of the $^{22}{\rm Ne}$ is much
longer than the evolutionary timescale for a DBV white dwarf. If possible,
we should include $^{22}{\rm Ne}$ as an adjustable parameter in future
fits to pulsation data. Independent observational tests of the core 
composition may also be possible through measurements of the rate of 
period change for white dwarf stars with well-isolated pulsation modes 
\citep[cf.][]{kep00,sul00}.

Although we have confirmed that the derived oxygen abundance does not
depend strongly on the sharpness of the He/C transition in the envelope,
we have not considered a fundamentally different form for the transition
zone. One possibility might be the form suggested by \cite{dk95}, who
included the effects of time-dependent diffusion in their white dwarf
models to try to establish a plausible evolutionary connection between the
PG~1159 stars and the DBVs. This led to two composition transitions in the
envelope: one from pure He to a He/C mixture at a fractional mass near
$10^{-6}$, and a second between the He/C mixture and the C/O core
occurring near $10^{-2.75}~m/M_*$. \cite{mnw00} found two families of
solutions corresponding approximately to these two layer masses in their
initial analysis of GD~358. This possibility merits further investigation,
but may not be sufficient in light of the recent detection of C in the
atmosphere of GD~358 \citep{pro00}, which cannot be explained by
traditional models of carbon dredge-up.

Finally, we could be assuming the wrong evolutionary origin for GD~358.
\cite{nrs81} suggested that some DB stars may be the end-product of a
merged interacting binary white dwarf like the AM CVn systems. The
internal oxygen abundances and profiles of such merged objects are
unlikely to be the same as their non-binary cousins, and we should be able
to distinguish them from one another through asteroseismology. This idea
was investigated observationally for AM CVn by \cite{pro95}, and some
theoretical consequences for DBV stars like GD~358 were quantified by
\cite{nw98}. But the latter authors conclude that if GD~358 has a binary
origin, it must have gone through an extremely long accretion
phase---approaching the theoretical limit.

The biggest problem at the moment is that we have only applied this global
model-fitting technique to one pulsating white dwarf, and the other DBV
stars have not been observed intensely enough and/or do not exhibit a
sufficient number of pulsation modes to allow a reliable analysis.  
\cite{hwn01} and the WET collaboration are currently attempting to remedy
this situation. Fortunately, each additional object will provide an
independent potential measurement of the \cago rate at stellar energies,
and white dwarfs with various masses should have slightly different
internal oxygen abundances---but all of them should be consistent with the
same underlying nuclear physics.

With the addition of an extra parameter for the H layer mass, it should be
possible to extend our method to the (more numerous) DAV white dwarfs, but
most of them also suffer from the problem of having too few observed
pulsation modes. \cite{kle98} attempted to overcome this problem by
assuming that the underlying mode structure in an individual DAV was
stable, and combining all of the frequencies visible in different
observing runs into a single list of modes. \cite{cle94} introduced the
concept of ensemble asteroseismology by assuming that the DAVs were an
approximately uniform class of objects that could be understood
collectively, and scaled the frequencies observed in many different DAVs
into a common asteroseismic proto-type. One or both of these techniques
may prove to be useful in our future modeling efforts.

\acknowledgements 

We owe special thanks to Mike Montgomery and the anonymous referee for 
helping us to understand the physical origin of our model sensitivity to 
the core composition. We thank Ed Nather, Steve Kawaler, and Joergen
Christensen-Dalsgaard for helpful discussions. This work was supported 
in part by the Danish National Research Foundation through its 
establishment of the Theoretical Astrophysics Center. Computational 
resources were provided by the University of Texas at Austin, Aarhus 
University, and by White Dwarf Research Corporation through grants from 
the Fund for Astrophysical Research and the American Astronomical Society.
Financial support was also provided through grant NAG5-9321 from NASA's 
Applied Information Systems Research Program.

\section*{APPENDIX}

The evolution code we used for the calculations presented here was derived
from WDEC, which was most recently described in detail by \cite{woo90}.
Descriptions of more recent updates to the constitutive physics can be
found in \cite{bra93} and \cite{mon98}, but for reference we provide an
overview of the details that are relevant to our DBV models. The equation
of state (EOS) for the cores of our models come from \cite{lam74}, and
from \cite{fgv77} for the envelopes. The C/O mixture in the interior and
the He/C mixture in the envelope are interpolated between pure
compositions using the additive volume technique \citep{fgv77}. We use an 
adaptation of
the method of \cite{af80} to describe the composition transition zones in
the envelope, as described in \S \ref{dcsec}. We use the updated OPAL
opacity tables from \cite{ir93}, neutrino rates from \cite{ito96}, and the
mixing-length prescription of \cite{bc71}. We use the Modified Ledoux
treatment of the Brunt-V\"ais\"al\"a frequency described by \cite{tfw90},
which includes the term that correctly accounts for composition gradients.

We determined the pulsation frequencies of our evolved models using the
adiabatic non-radial oscillation code described by \cite{kaw86}, which
solves the pulsation equations using the Runge-Kutta-Fehlberg method. The
periods resulting from the adiabatic approximation typically differ from
the non-adiabatic results by only a few thousandths of a second, which is
well below the present level of observational noise.

The validity of our DB white dwarf evolution/pulsation models have been
investigated by \cite{ba97}, who used their independent state-of-the-art
code to verify conclusions made by \cite{bw94a}, which were based on
the same code we have used in our study. \citeauthor{ba97} include a more
recent convective theory in their models that naturally leads to an
efficiency between ML2 and ML3. This difference does not matter in the
context of our paper, since we have demonstrated in \S \ref{mlsec} the
weak dependence of our final result on the convective prescription.

\end{document}